\documentclass[12pt,titlepage]{article}
\pdfoutput=1 

\usepackage{amsmath}
\usepackage{amssymb}
\usepackage{caption}
\usepackage{authblk}
\usepackage{subcaption}

\usepackage{multirow}
\usepackage{graphicx}
\usepackage{enumerate}
\usepackage[hdivide={2cm,*,2cm},vdivide={1in,*,1in}]{geometry}
\usepackage{setspace}
\usepackage[retainorgcmds]{IEEEtrantools}
\usepackage{natbib}
\usepackage{color}
\newcommand{\ud}{\, {\rm d} \kern-.015em }

\title{Adaptive Survival Trials}
\author[1]{Dominic Magirr}
\author[2]{Thomas Jaki}
\author[1]{Franz Koenig}
\author[1]{Martin Posch}
\affil[1]{Section for Medical Statistics, Medical University of Vienna, Austria}
\affil[2]{Department of Mathematics and Statistics, Lancaster University, UK}

\begin{document} 

\maketitle

\begin{abstract}
Mid-study design modifications are becoming increasingly accepted in confirmatory clinical trials, so long as appropriate methods are applied such that error rates are controlled. It is therefore unfortunate that the important case of time-to-event endpoints is not easily handled by the standard theory. We analyze current methods that allow design modifications to be based on the full interim data, i.e., not only the observed event times but also secondary endpoint and safety data from patients who are yet to have an event. We show that the final test statistic may ignore a substantial subset of the observed event times. Since it is the data corresponding to the earliest recruited patients that is ignored, this neglect becomes egregious when there is specific interest in learning about long-term survival. An alternative test  incorporating all event times is proposed, where a conservative assumption is made in order to guarantee type I error control. We examine the properties of our proposed approach using the example of a clinical trial comparing two cancer therapies. 

\end{abstract}

\section{Introduction}

There are often strong ethical and economic arguments for conducting interim analyses of an ongoing clinical trial and for making changes to the design if warranted by the accumulating data. One may decide, for example, to increase the sample size on the basis of promising interim results. Or perhaps one might wish to drop a treatment from a multi-arm study on the basis of unsatisfactory safety data. Owing to the complexity of clinical drug development, it is not always possible to anticipate the need for such modifications, and therefore not all contingencies can be dealt with in the statistical design.

Unforeseen interim modifications complicate the (frequentist) statistical analysis of the trial considerably. Over recent decades many authors have investigated so-called ``adaptive designs'' in an effort to maintain the concept of type I error control \citep{bauer94,proschan95,muller01,hommel01}. Although Bayesian adaptive methods are becoming increasing popular, type I error control is still deemed important in the setting of a confirmatory phase III  trial \citep[p. 6]{berry10}, and  recent years have seen  hybrid adaptive designs proposed, whereby the interim decision is based on Bayesian methods, but the final hypothesis test remains frequentist \citep{brannath2009confirmatory,di2011time}.

 While the  theory of adaptive designs is now well understood if responses are observed immediately, subtle problems arise when responses are delayed, e.g., in survival trials. 

\cite{schafer01} proposed adaptive survival tests that are constructed using the independent increments property of logrank test statistics  \citep[c.f.,][]{wassmer06,desseaux2007flexible,  jahn09}.  However, as pointed out by \cite{bauer04}, these methods only work if interim decision making is based solely on the interim logrank test statistics  and any secondary endpoint data from patients  who have already had an event. In other words, investigators must remain blind to the data from patients who are censored at the interim analysis. \cite{irle12} argue that decisions regarding interim design modifications should be as substantiated as possible, and  propose a test procedure that allows investigators to use the full interim data. This methodology, similar to that of \cite{jenkins10}, does not require any assumptions regarding the joint distribution of survival times and short-term secondary endpoints, as do, e.g.,  the methods proposed by \cite{stallard2010confirmatory}, \cite{friede2011designing, friede2012conditional} and \cite{hampson2013group}.

The first goal of this article is to clarify the proposals of  \cite{jenkins10} and \cite{irle12}, showing that they are both based on weighted inverse-normal test statistics \citep{lehmacher99}, with the common disadvantage that the final test statistic may ignore a substantial subset of the observed survival times. This is a serious limitation, as disregarding part of the observed data is generally considered inappropriate even if statistical error probabilities are controlled -- see, for example, the discussion on overrunning in group sequential trials \citep{hampson2013group}. Our secondary goal is therefore to propose an alternative test that retains the strict type I error control and flexibility of the aforementioned designs, but bases the final test decision on a statistic that takes into account all available survival times.  As ever, there is no  free lunch, and the assumption that we  require to ensure type I error control induces a certain amount of conservatism. We evaluate the properties of our proposed approach using the example of a clinical trial comparing two cancer therapies.

\section{Adaptive Designs \label{secStandardAD}}
\subsection{Standard theory}
A comprehensive account of adaptive design methodology can be found in \cite{bretz09}. For testing a null hypothesis, $H_0:\theta = 0$, against the one-sided alternative, $H_a:\theta>0$, the archetypal two-stage adaptive test statistic  is of the form $f_1(p_1) + f_2(p_2)$,
where $p_1$ is the p-value based on the first-stage data, $p_2$ is the p-value from the (possibly adapted) second-stage test, and $f_1$ and $f_2$ are prespecified monotonically decreasing functions. Consider the simplest case that no early rejection of the null hypothesis is possible at the end of the first stage. The null hypothesis is rejected at level $\alpha$ whenever $f_1(p_1)+f_2(p_2)>k$, where $k$ satisfies
\[
\int_0^1 \int_0^1 \mathbf{1}\left\lbrace f_1(p_1)+f_2(p_2) \leq k \right\rbrace \ud p_1 \ud p_2=1-\alpha. 
\]

In their seminal paper, \cite{bauer94} took $f_i(p_i)=-\log(p_i)$ for $i=1,2$. We will restrict attention to the weighted inverse-normal test statistic \citep{lehmacher99},
\begin{equation}\label{eqnStandardTS} 
Z=w_1\Phi^{-1}(1-p_1)+w_2\Phi^{-1}(1-p_2),
\end{equation} 
where $\Phi$ denotes the standard normal distribution function and $w_1$ and $w_2$ are prespecified weights such that $w_1^2+w_2^2=1$.  If $Z>\Phi^{-1}(1-\alpha)$, then $H_0$ may be rejected at level $\alpha$. The assumptions required to make this a valid level-$\alpha$ test are as follows \cite[see][]{brannath12}.

\textbf{Assumption 1}


Let $X_1^{\text{int}}$ denote the data available at the interim analysis, where $X_1^{\text{int}} \in \mathbb{R}^n$ with distribution function $G(x_1^{\text{int}};\theta)$. The calendar time of the interim analysis will be denoted $T^{\text{int}}$. In general, $X_1^{\text{int}}$ will contain information not only concerning the primary endpoint, but also measurements on secondary endpoints and safety data. It is assumed that the first-stage p-value function $p_1:\mathbb{R}^n\rightarrow \left[0,1\right]$ satisfies $$\int_{\mathbb{R}^n}\mathbf{1}{\left\lbrace p_1 (x_1^{\text{int}})\leq u \right\rbrace}\ud G(x_1^{\text{int}};0)\leq u \text{ for all }u \in \left[ 0,1\right].$$

\textbf{Assumption 2}

At the interim analysis, a second-stage design $d$ is chosen. The second-stage design is allowed to depend on the unblinded first-stage data without prespecifying an adaptation rule. Denote the second-stage data by $Y$, where $Y \in \mathbb{R}^m$. It is assumed that the distribution function of $Y$, denoted by $F_{d,x_1^{\text{int}}}(y,\theta)$, is known for all possible second stage designs, $d$, and all first-stage outcomes, $x_1^{\text{int}}$.

\textbf{Assumption 3}

The second-stage p-value function $p_2:\mathbb{R}^m\rightarrow \left[ 0,1 \right]$ satisfies $\int_{\mathbb{R}^m}\mathbf{1}\left\lbrace p_2 (y)\leq u \right\rbrace\ud F_{d,x_1^{\text{int}}}(y;0)\leq u \text{ for all }u \in \left[ 0,1\right]$.

\subsection{Immediate responses}

\begin{figure}[h]
\begin{center}
\includegraphics[width=0.7\textwidth,trim= 0cm 20cm 5cm 0cm]{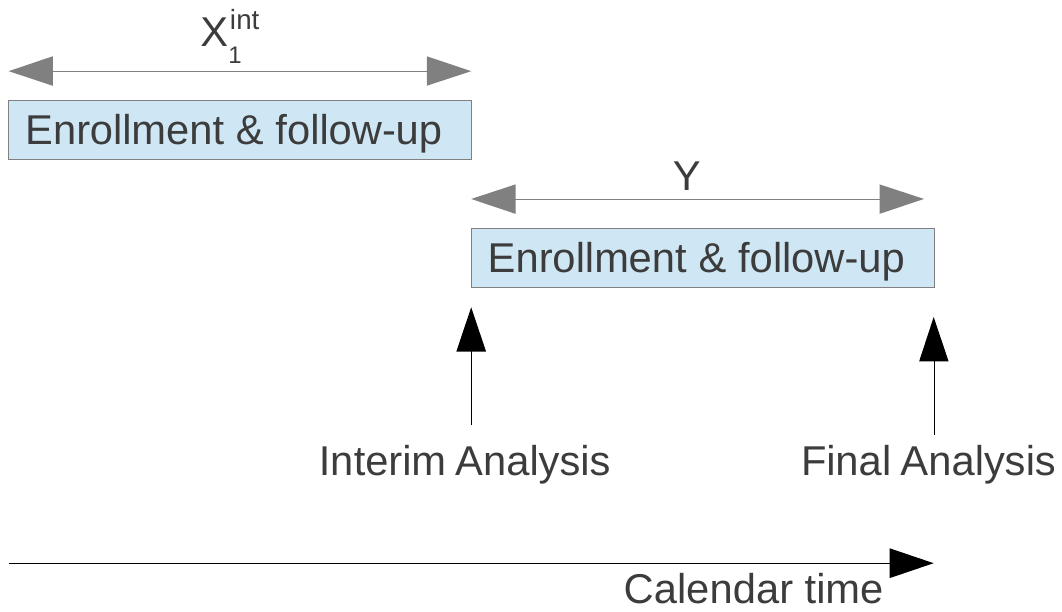}
\end{center}
\caption{\label{figStandardAD} Schematic of a standard two-stage adaptive trial with immediate response.}
\end{figure}

The aforementioned assumptions are easy to justify when primary endpoint responses are observed more-or-less immediately. In this case  $X_1^{\text{int}}$ contains the responses of all patients recruited prior to the interim analysis. A second-stage design $d$ can subsequently be chosen with the responses from a new cohort of patients contributing to $Y$ (Figure \ref{figStandardAD}).

\subsection{Delayed responses and the independent increments assumption\label{secII}}

\begin{figure}[h]
\begin{center}
\includegraphics[width=0.8\textwidth,trim= 0cm 20cm 3cm 0cm]{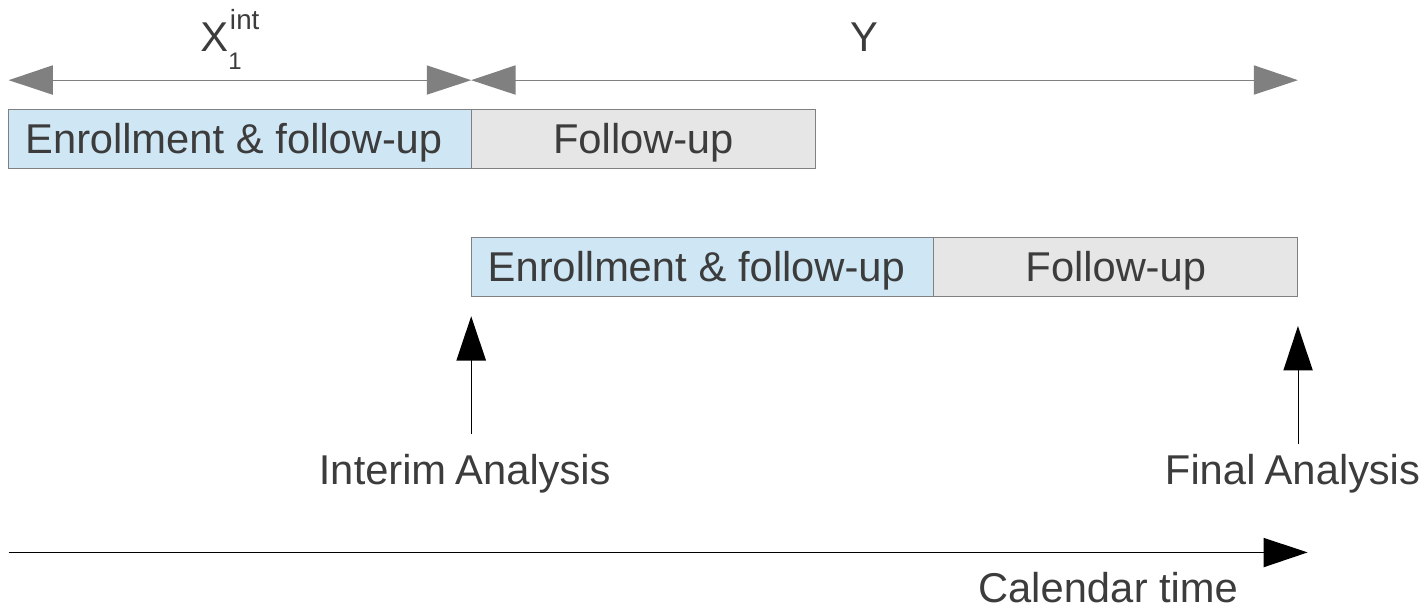}
\end{center}
\caption{\label{figXandY2} Schematic of a two-stage adaptive trial with delayed response under the independent increments assumption.}
\end{figure}

 An interim analysis may take place whilst some patients have entered the study but have yet to provide a data point on the primary outcome measure. Most approaches to this problem \citep[e.g.,][]{schafer01,wassmer06,jahn09} attempt to take advantage of the well known independent increments structure of score statistics in group sequential designs \citep{jennison00}. As pictured in Figure \ref{figXandY2}, $X_1^{\text{int}}$ will generally include responses on short-term secondary endpoints and safety data from patients who are yet to provide a primary outcome measure, while $Y$ consists of some delayed responses from patients recruited prior to $T^{\text{int}}$, mixed together with responses from a new cohort of patients.

Let $S({X_1^{\text{int}}})$ and $\mathcal{I}({X_1^{\text{int}}})$ denote the score statistic and Fisher's information for $\theta$, calculated from primary endpoint responses in $X_1^{\text{int}}$. Assuming suitable regularity conditions, the asymptotic null distribution of $S({X_1^{\text{int}}})$ is Gaussian with mean zero and variance $\mathcal{I}({X_1^{\text{int}}})$ \citep[][p. 107]{cox1979theoretical}. The independent increments assumption is that for all first-stage outcomes $x_1^{\text{int}}$ and second-stage designs $d$, the null distribution of $Y$ is such that
\begin{equation}\label{eqnIndInc}
S(x_1^{\text{int}},Y)-S(x_1^{\text{int}}) \sim \mathcal{N}\left\lbrace 0,\mathcal{I}(x_1^{\text{int}},Y)-\mathcal{I}(x_1^{\text{int}})\right\rbrace,
\end{equation}
at least approximately, where $S_{X_1^{\text{int}},Y}$ and $\mathcal{I}_{X_1^{\text{int}},Y}$ denote the score statistic and Fisher's information for $\theta$, calculated from primary endpoint responses in $(X_1^{\text{int}},Y)$.

Unfortunately, (\ref{eqnIndInc}) is seldom realistic in an adaptive setting. \cite{bauer04} show that if the adaptive strategy at the interim analysis is dependent on short-term outcomes in $X_1^{\text{int}}$ that are correlated with primary endpoint outcomes in $Y$, i.e., from the same patient, then a naive appeal to the independent increments assumption can lead to very large type I error inflation.

\subsection{Delayed responses with ``patient-wise separation''\label{secPatSplit}}

An alternative approach, which we shall coin ``patient-wise separation'', redefines the first-stage p-value, $p_1: \mathbb{R}^p\rightarrow \left[0,1 \right]$, to be a function of $X_1$,  where $X_1$ denotes all the data from patients recruited prior to $T^{\text{int}}$, followed-up until calendar time $T^{\max}$ -- which corresponds to the prefixed maximum duration of the trial. It is assumed that $X_1$ takes values in $\mathbb{R}^p$ according to distribution function  $\tilde{G}(x_1;\theta)$. Assumption \textbf{1} is replaced with:\\
\begin{equation}\label{eqnA1Star}
\int_{\mathbb{R}^p}\mathbf{1}{\left\lbrace p_1 (x_1)\leq u \right\rbrace}\ud \tilde{G}(x_1;0)\leq u \text{ for all }u \in \left[ 0,1\right].
\end{equation}
In this case $p_1$ may not be observable at the time the second-stage design $d$ is chosen. This is not a problem, as long as no early rejection at the end of the first stage is foreseen. Any interim decisions, such as increasing the sample size, do not require any knowledge of $p_1$. It is assumed that $Y$ consists of responses from a new cohort of patients, such that $x_1^{\text{int}}$ could be formally replaced with $x_1$ in assumptions \textbf{2} and \textbf{3}. We call this ``patient-wise separation'' because data from the same patient cannot contribute to both $p_1$ and $p_2$.

\cite{liu05} consider such an approach for a clinical trial  where a patient's primary outcome is  measured  after a fixed period of follow-up, e.g., 4 months. Provided that one is willing to wait for all responses, it is straightforward to prespecify a first-stage p-value function such that (\ref{eqnA1Star}) holds.

For an adaptive trial with a time-to-event endpoint, however, one must be very careful to ensure that (\ref{eqnA1Star}) holds, as one is typically not prepared to wait for all first-stage patients -- those patients recruited prior to $T^{\text{int}}$ -- to have an event. Rather, $p_1$ is defined as the p-value from an, e.g., logrank test applied to the data from first-stage patients followed up until time $T_1$, for some $T_1<T^{\max}$. In this case it is vital that $T_1$ be fixed at the start of the trial, either explicitly or implicitly \citep{jenkins10,irle12}. Otherwise, if $T_1$ were to depend on the adaptive strategy at the interim analysis, this would impact the distribution of $p_1$ and could lead to type I error inflation.

\begin{figure}[h]
\begin{center}
\includegraphics[width=0.8\textwidth,trim= 0cm 20cm 3cm 0cm]{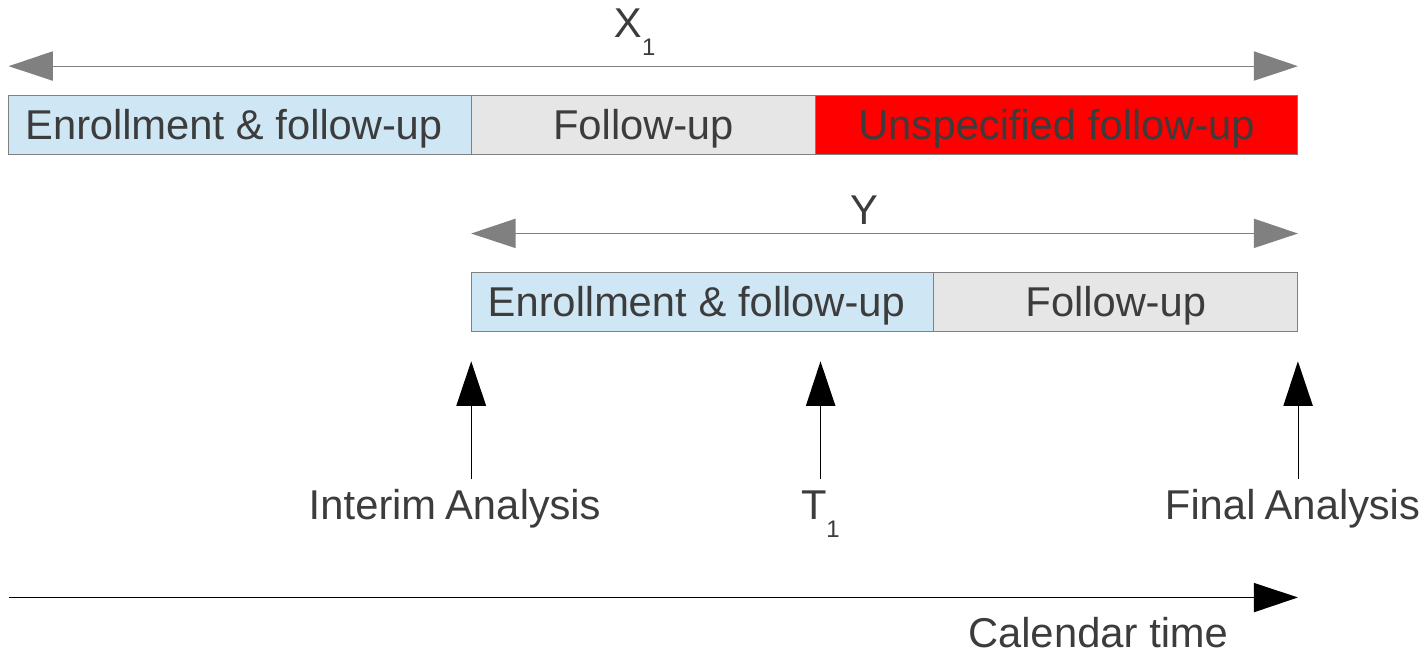}
\end{center}
\caption{\label{figDRADwithOS} Schematic of a two-stage adaptive trial with ``patient-wise separation''.}
\end{figure}

The situation is represented pictorially in Figure \ref{figDRADwithOS}. An unfortunate consequence of prefixing $T_1$ is that this will not, in all likelihood, correspond to the end of follow-up for second-stage patients. All events of first-stage patients that occur after $T_1$ make no contribution to the statistic (\ref{eqnStandardTS}); they are ``thrown away''.

\section{Adaptive Survival Studies \label{secAdSurv}}
\subsection{\cite{jenkins10} method \label{secJenkins}}

 Consider a randomized clinical trial comparing survival times on  an experimental treatment, $E$, with those on a  control treatment, $C$. We will focus on the logrank statistic for testing the null hypothesis $H_0:\theta=0$ against the one-sided alternative $H_a:\theta>0$, where $\theta$ is the log hazard ratio, assuming proportional hazards. Let $D_{1}(t)$ and $S_{1}(t)$ denote the number of uncensored events and the usual  logrank score statistic, respectively, based on the data from first-stage patients -- those patients recruited prior to the interim analysis --  followed up until calendar time $t$, $t \in \left[0,T^{\max}\right]$. Under the null hypothesis, assuming equal allocation and a large number of events, the variance of $S_{1}(t)$ is approximately equal to $D_{1}(t)/4$ \citep[e.g.,][Section 3.4]{whitehead97}. The first-stage p-value must be calculated at a prefixed time point $T_{1}$:
\begin{equation}\label{eqnP1}
p_1=1-\Phi\left[ 2\left\lbrace S_{1}(T_{1})\right\rbrace /\left\lbrace D_{1}(T_{1})\right\rbrace^{1/2}\right].
\end{equation}
There are two possible ways of specifying $T_{1}$ in (\ref{eqnP1}). From a practical perspective, a \textit{calendar time} approach is often attractive as $T_{1}$ is  specified explicitly, which facilitates straightforward planning. On the other hand, this can produce a misspowered study if the recruitment rate and/or survival times differ markedly from those anticipated. An \textit{event driven} approach may be preferred, whereby the number of  events is prefixed at $d_{1}$, say, and 
\begin{equation}\label{eqnTFix}
T_{1}:=\min\left\lbrace t :D_{1}(t)=d_{1}\right\rbrace.
\end{equation}

\cite{jenkins10} describe a ``patient-wise separation'' adaptive survival trial, with test statistic (\ref{eqnStandardTS}), first-stage p-value (\ref{eqnP1}) and $T_{1}$ defined as in (\ref{eqnTFix}). While their focus is on subgroup selection, we will appropriate their method for the simpler situation of a single comparison, where at the interim analysis one has the possibility to alter the pre-planned number of events from second-stage patients -- i.e., those patients recruited post $T^{\text{int}}$. All that remains to be specified at the design stage is the choice of weights $w_1$ and $w_2$. It is anticipated that $p_2$ will be the p-value corresponding to a logrank test based on second-stage patients, i.e., 
\[
p_2=1-\Phi\left[2 S_2(T_2^*)/\left\lbrace D_2(T_2^*)\right\rbrace^{1/2}\right],
\]
where $T_2^*:=\min \left\lbrace t:D_2(t)=d_2^*\right\rbrace$ with $S_2(t)$ and $D_2(t)$ defined analogously to $S_1(t)$ and $D_1(t)$, and $d_2^*$ is to be specified at the interim analysis.
 Ideally, the weights should be chosen in proportion to the information (number of events) contributed from each stage. In an adaptive trial, it is impossible to achieve the correct weighting in every scenario. \citeauthor{jenkins10} prespecify the \textit{envisioned} number of second-stage events, $d_{2}$, and choose weights $w_1=\left\lbrace d_{1}/(d_{1}+d_{2})\right\rbrace^{1/2}$ and $w_2=\left\lbrace d_{2}/(d_{1}+d_{2})\right\rbrace^{1/2}$.

\subsection{\cite{irle12} method \label{secIrle}}

\cite{irle12} propose an alternative procedure.   Instead of  explicitly combining stage-wise p-values, they employ the closely related  \textit{conditional error} approach \citep{proschan95,posch99,muller01}. 

They begin by prespecifying a level-$\alpha$ test with decision function, $\varphi$, taking values in $\left\lbrace 0,1\right\rbrace$ corresponding to nonrejection and rejection of $H_0$, respectively. For a survival trial, this entails specifying the sample size, duration of follow-up, test statistic, recruitment rate, etc. Then, at some (not necessarily prespecified) timepoint, $T^{\text{int}}$, an interim analysis is performed. The timing of the interim analysis induces a partition of the trial data, $(X_1,X_2)$, where $X_1$ and $X_2$ denote the data from patients recruited prior- $T^{\text{int}}$ and post- $T^{\text{int}}$, respectively,  followed-up until time $T^{\max}$. More specifically, \cite{irle12} suggest the decision function
\begin{equation}\label{eqnVarPhi}
\varphi(X_1,X_2)=\mathbf{1}\left[ 2 S_{1,2}(T_{1,2})/\left\lbrace D_{1,2}(T_{1,2})\right\rbrace^{1/2}>\Phi^{-1}(1-\alpha)\right],
\end{equation}
where  $D_{1,2}(T_{1,2})$ and $S_{1,2}(T_{1,2})$ denote the number of uncensored events and the usual logrank score statistic, respectively, based on data from all patients (from both stages) followed-up until time $T_{1,2}$, where $T_{1,2}:=\min\left\lbrace t :D_{1,2}(t)=d_{1,2}\right\rbrace$ for some prespecified number of events $d_{1,2}$.

At the interim analysis, the general idea is to use the unblinded first-stage data $x_1^{\text{int}}$ to define a second-stage design, $d$, without the need for a prespecified adaptation strategy. Again, the definition of $d$ includes factors such as sample size, follow-up period, recruitment rate, etc., in addition to a second-stage decision function $\psi_{x_1^{\text{int}}}: \mathbb{R}^m \rightarrow \left\lbrace 0,1 \right\rbrace$ based on second-stage data $Y \in \mathbb{R}^m$. \cite{irle12} focus their attention on a specific design change; namely, the possibility of increasing the number of events from $d_{1,2}$ to $d_{1,2}^*$ by extending the follow-up period. They assume that $Y:=(X_1,X_2)\setminus X_1^{\text{int}}$ and propose the second-stage decision function
\begin{equation}\label{eqnPsi}
\psi_{x_1^{\text{int}}}(Y)=\mathbf{1}\left[2 S_{1,2}(T_{1,2}^*)/\left\lbrace D_{1,2}(T_{1,2}^*)\right\rbrace^{1/2}\geq b^*\right],
\end{equation}
where $T^*_{1,2}:=\min\left\lbrace t :D_{1,2}(t)=d^*_{1,2}\right\rbrace$ and $b^*$ is a cutoff value that must be determined. Ideally, one would like to choose $b^*$ such that $E_{H_0}(\psi_{X_1^{\text{int}}}\mid X_1^{\text{int}} = x_1^{\text{int}})=E_{H_0}(\varphi \mid X_1^{\text{int}} = x_1^{\text{int}})$, as this would ensure that
\begin{IEEEeqnarray}{CCCCCCCCC }\label{eqnCEalpha}
E_{H_0}(\psi_{X_1^{\text{int}}}) & = &E_{H_0}\left\lbrace E_{H_0} \left( \psi_{X_1^{\text{int}}} \mid X_1^{\text{int}}\right) \right\rbrace & = & E_{H_0}\left\lbrace E_{H_0} \left( \varphi \mid X_1^{\text{int}}\right) \right\rbrace 
 & = & E_{H_0}(\varphi) 
 & = & \alpha,
\end{IEEEeqnarray}
i.e., the overall procedure controls the type I error rate at level $\alpha$. Unfortunately, this approach is not directly applicable in a survival trial where $X_1^{\text{int}}$ contains short-term data from first-stage patients surviving beyond $T^{\text{int}}$. This is because it is impossible to calculate $E_{H_0}(\varphi \mid X_1^{\text{int}} = x_1^{\text{int}})$ and $E_{H_0}(\psi_{X_1^{\text{int}}}\mid X_1^{\text{int}} = x_1^{\text{int}})$, owing to the unknown joint distribution of survival times and the secondary/safety endpoints already observed at the interim analysis, c.f. Section \ref{secII}. \cite{irle12} get around this problem by conditioning on additional variables; namely, $S_{1}(T_{1,2})$ and $S_1(T_{1,2}^*)$. 
Choosing $\psi_{x_1^{\text{int}}}$ such that 
\begin{equation*}
E_{H_0}\left\lbrace \psi_{X_1^{\text{int}}} \mid X_1^{\text{int}} = x_1^{\text{int}}, S_1(T_{1,2})=s_1, S_1(T_{1,2}^*)=s_1^*\right\rbrace=E_{H_0}\left\lbrace \varphi \mid X_1^{\text{int}} = x_1^{\text{int}},  S_1(T_{1,2})=s_1, S_1(T_{1,2}^*)=s_1^*\right\rbrace
\end{equation*}
 ensures that $E_{H_0}(\psi_{X_1^{\text{int}}})=\alpha$ following the same argument as (\ref{eqnCEalpha}).

\cite{irle12} show that, asymptotically, $$E_{H_0}\left\lbrace \varphi \mid X_1^{\text{int}} = x_1^{\text{int}},S_1(T_{1,2})=s_1, S_1(T_{1,2}^*)=s_1^*\right\rbrace =E_{H_0}\left\lbrace \varphi \mid S_1(T_{1,2})=s_1 \right\rbrace$$ and $$E_{H_0}\left\lbrace \psi_{X_1^{\text{int}}} \mid X_1^{\text{int}} = x_1^{\text{int}},S_1(T_{1,2})=s_1, S_1(T_{1,2}^*)=s_1^*\right\rbrace=E_{H_0}\left\lbrace \psi_{x_1^{\text{int}}} \mid S_1(T_{1,2}^*)=s_1^* \right\rbrace. $$ In each case, calculation of the right-hand-side  is facilitated by the asymptotic result that, assuming equal allocation under the null hypothesis,
\begin{equation}\label{eqnDist}
\left( \begin{array}{c}S_{1}(t) \\ S_{1,2}(t) -S_{1}(t)\end{array}\right) \sim \mathcal{N}\left(\left(\begin{array}{c}0\\0\end{array}\right),\left(\begin{array}{c c}D_{1}(t)/4 & 0 \\ 0 &  \left\lbrace D_{1,2}(t)-D_{1}(t)\right\rbrace/4 \end{array} \right)\right),
\end{equation}
for $t \in \left[ 0, T \right]$, where $T$ is sufficiently large such that all events of interest occur prior to $T$.

One remaining subtlety is that  $E_{H_0}\left\lbrace \psi_{x_1^{\text{int}}} \mid S_1(T_{1,2}^*)=s_1\right\rbrace$ can only calculated at calendar time $T_{1,2}^*$, where $T_{1,2}^*>T^{\text{int}}$. Determination of $b^*$ must therefore be postponed until this later time.

It is shown in Appendix A that $\psi_{X_1^{\text{int}}}=1$ if and only if $Z>\Phi^{-1}(1-\alpha)$, where $Z$ is defined as in (\ref{eqnStandardTS}) with $p_1$ defined as in (\ref{eqnP1}), $T_1$ defined as equal to $T_{1,2}$, the second-stage p-value function defined as 
\begin{equation}\label{eqnP2X+}
p_2(Y)=1-\Phi\left[ 2\left\lbrace S_{1,2}(T^*_{1,2})-S_{1}(T^*_{1,2})\right\rbrace / \left\lbrace d^*_{1,2}-D_{1}(T^*_{1,2})\right\rbrace^{1/2} \right],
\end{equation}
and the specific choice of weights:
\begin{equation}\label{eqnIrleWeights}
 w_1=\left\lbrace D_{1}(T_{1,2})/ d_{1,2} \right\rbrace^{1/2}\text{ and }w_2=\left[\left\lbrace d_{1,2}- D_{1}(T_{1,2})\right\rbrace / d_{1,2}\right]^{1/2}.
\end{equation}


\textbf{Remark 1.}  In a sense, the \citeauthor{irle12} method can be thought of as a special case of the \citeauthor{jenkins10} method, with a clever way of implicitly defining the weights and the end of first-stage follow-up, $T_1$. It has two potential advantages. Firstly, the timing of the interim analysis need not be prespecified -- in theory, one is permitted to monitor the accumulating data and at any moment decide that design changes are necessary. Secondly, if no changes to the design are necessary, i.e., the trial completes as planned at calendar time $T_{1,2}$, then the original test (\ref{eqnVarPhi}) is performed. In this special case, no data is ``thrown away''.

\textbf{Remark 2.} From first glance at (\ref{eqnPsi}), it may appear that the  data from first-stage patients, accumulating after $T_{1,2}$, is never ``thrown away''. However, this data is still effectively ignored. We have shown that the procedure is equivalent to a p-value combination approach where $p_1$ depends only on data available at time $T_1:=T_{1,2}$. In addition, the distribution of $p_2$ is asymptotically independent of the data from first-stage patients: note that  $S_{1,2}(T^*_{1,2})-S_{1}(T^*_{1,2})$ and $S_{2}(T^*_{1,2})$ are asymptotically equivalent \cite[][remark 1]{irle12}. The procedure therefore fits our description of a ``patient-wise separation'' design, c.f. Section \ref{secPatSplit}, and the  picture is the same as in Figure \ref{figDRADwithOS}. The first-stage patients have in effect been censored at $T_{1,2}$, despite having been followed-up for longer.
 
This fact has important implications for the choice of $d_{1,2}^*$. If one chooses $d_{1,2}^*$ based on conditional power arguments, one should be aware that the effective sample size has not increased by $d_{1,2}^*-d_{1,2}$. Rather, it has increased by $d_{1,2}^*-d_{1,2}-\left\lbrace D_1(T_{1,2}^*)-D_1(T_{1,2})\right\rbrace$, which could be very much smaller.

\textbf{Remark 3.}  A potential disadvantage of the \cite{irle12} method  is that it is not possible to decrease the number of events  (nor decrease the recruitment rate) at the interim analysis, as one must observe at least $d_{1,2}$ events (in the manner specified by the original design) to be able to calculate the conditional error probability $E_{H_0}\left\lbrace \varphi \mid S_1(T_{1,2})\right\rbrace$.  

\begin{figure}[h]
\begin{center}
\includegraphics[width=0.8\textwidth,trim= 0cm 18cm 3cm 0cm]{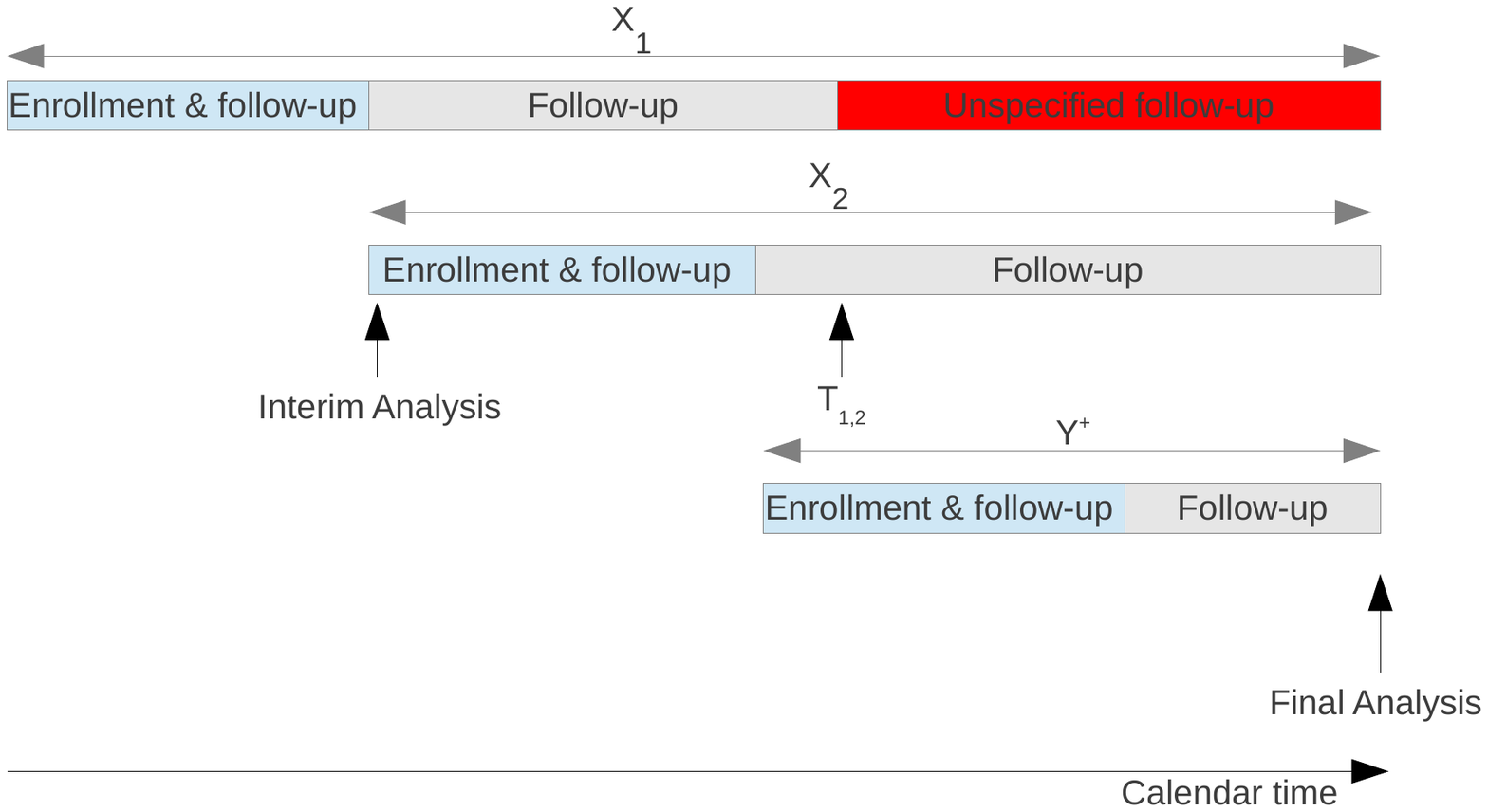}
\end{center}\caption{\label{figIS} Extension of the \citeauthor{irle12} approach to allow a prolonged recruitment period.}
\end{figure}

In addition, one is not permitted to increase the recruitment rate following the interim analysis, nor to prolong the recruitment period beyond that prespecified by the original design. In order to allow such design changes, a small extension is necessary. While the conditional error probability remains $E_{H_0}\left\lbrace \varphi \mid S_1(T_{1,2})\right\rbrace$, the second-stage data must be split into two parts, $Y=\left\lbrace (X_1,X_2)\setminus X_1^{\text{int}},Y^+ \right\rbrace$, where $Y^+$ consists of responses from an additional cohort of patients, not specified by the original design (see Figure \ref{figIS}). The second-stage test (\ref{eqnPsi}) can be replaced with, e.g.,
\begin{equation*}
\psi_{x_1^{\text{int}}}(Y)=\mathbf{1}\left[2 S_{2,+}(T_{2,+})/\left\lbrace D_{2,+}(T_{2,+})\right\rbrace^{1/2}\geq b^*\right],
\end{equation*}
where $D_{2,+}(T_{2,+})$ and $S_{2,+}(T_{2,+})$ are the observed number of events and the usual logrank score statistic, respectively, based on the responses of all patients recruited post $T^{\text{int}}$, and $T_{2,+}:=\min\left\lbrace t :D_{2,+}(t)=d_{2,+}\right\rbrace$ for some $d_{2,+}$ defined at time $T^{\text{int}}$. Again, determination of $b^*$ must be postponed until time $\max (T_{1,2},T_{2,+})$.

\subsection{Effect of unspecified follow-up data \label{secUnspecified}}

Continuing with the set up and notation of Section \ref{secJenkins} (which we have shown also fits the \cite{irle12} method), the adaptive test statistic is 
\begin{IEEEeqnarray}{rCl}\label{eqnAgain}
Z & = & w_1 \Phi^{-1}(1-p_1) + w_2 \Phi^{-1}(1-p_2)\nonumber\\
& = & 2w_1 S_{1}(T_{1})/D_{1}(T_{1})^{1/2}  + w_2 \Phi^{-1}(1-p_2). 
\end{IEEEeqnarray}
Suppose, however, that the trial continues until calendar time $T^*$, where $T^*>T_{1}$. Strictly speaking, the  data from first-stage patients -- those patients recruited prior to $T^{\text{int}}$ --  accumulating between times $T_{1}$ and $T^*$ should be ``thrown away''. In this section we will investigate what happens, in a worst case scenario, if this illegitimate data is naively incorporated into $Z$. Specifically, we find the maximum type I error associated with the test statistic
\begin{equation}\label{eqnZStar}
Z^*=2w_1 S_{1}(T^*)/D_{1}(T^*)^{1/2} + w_2 \Phi^{-1}(1-p_2).
\end{equation}

Since in practice $T^*$ depends on the interim data in a complicated way, the null distribution of (\ref{eqnZStar}) is unknown. One can, however, consider properties of the stochastic process
\begin{equation*}\label{eqnStocProc}
Z(t)=2w_1S_{1}(t)/D_1(t)^{1/2}+ w_2 \Phi^{-1}(1-p_2),\qquad t \in \left[T_1,T^{\max}\right].
\end{equation*}
In other words, we consider continuous monitoring of the logrank statistic based on first-stage patient data. The worst-case scenario assumption is that the responses on short-term secondary endpoints, available at the interim analysis, can be used to predict the exact calendar time the process $Z(t)$ reaches its maximum. In this case, one could attempt to engineer the second stage design such that $T^*$ coincides with this timepoint, and the worst-case type I error rate is therefore
\begin{equation}\label{findProb}
P_{H_0}\left\lbrace \max_{t\geq T_1}Z(t) > \Phi^{-1}(1-\alpha)\right\rbrace.
\end{equation}

Although the worst-case scenario assumption is clearly unrealistic, (\ref{findProb}) serves as an upper bound on the type I error rate. It can be found approximately via standard Brownian motion results. Define the \textit{information time} at calendar time $t$ to be $u = D_1(t)/D_1(T^{\max})$, and let $S_1(u)$ denote the logrank score statistic based on first-stage patients, followed-up until information time $u$. It can be shown that $B(u):=2S_1(u)/\left\lbrace D_1(T^{\max})\right\rbrace^{1/2}$ behaves asymptotically like a Brownian motion with drift $\xi:=\theta\left\lbrace D_1(T^{\max})/4\right\rbrace^{1/2}$ \citep[][p. 101]{proschan06}.


We wish to calculate
\begin{equation}\label{eqnWienWeights}
P_{\theta=0}\left\lbrace \max_{t\geq T_1} Z(t) > \Phi^{-1}(1-\alpha)\right\rbrace = \int_{0}^{1}P_{\theta=0}\left[ \max_{u=u_1}^{1} B(u) > u^{1/2}w_1^{-1}\left\lbrace \Phi^{-1}(1-\alpha)-w_2\Phi^{-1}(1- p_2)\right\rbrace\right] \ud p_2,
\end{equation}
where $u_1=D_1(T_1)/D_1(T^{\max})$. While the integrand on the right-hand-side is difficult to evaluate exactly, it can be found to any required degree of accuracy by replacing the square root stopping  boundary with a piecewise linear boundary \citep{wang97}. Some further details are provided in Appendix B.

\begin{figure}[h]
\begin{center}
\includegraphics[width=0.7\textwidth,trim= 0cm 0cm 0cm 0cm]{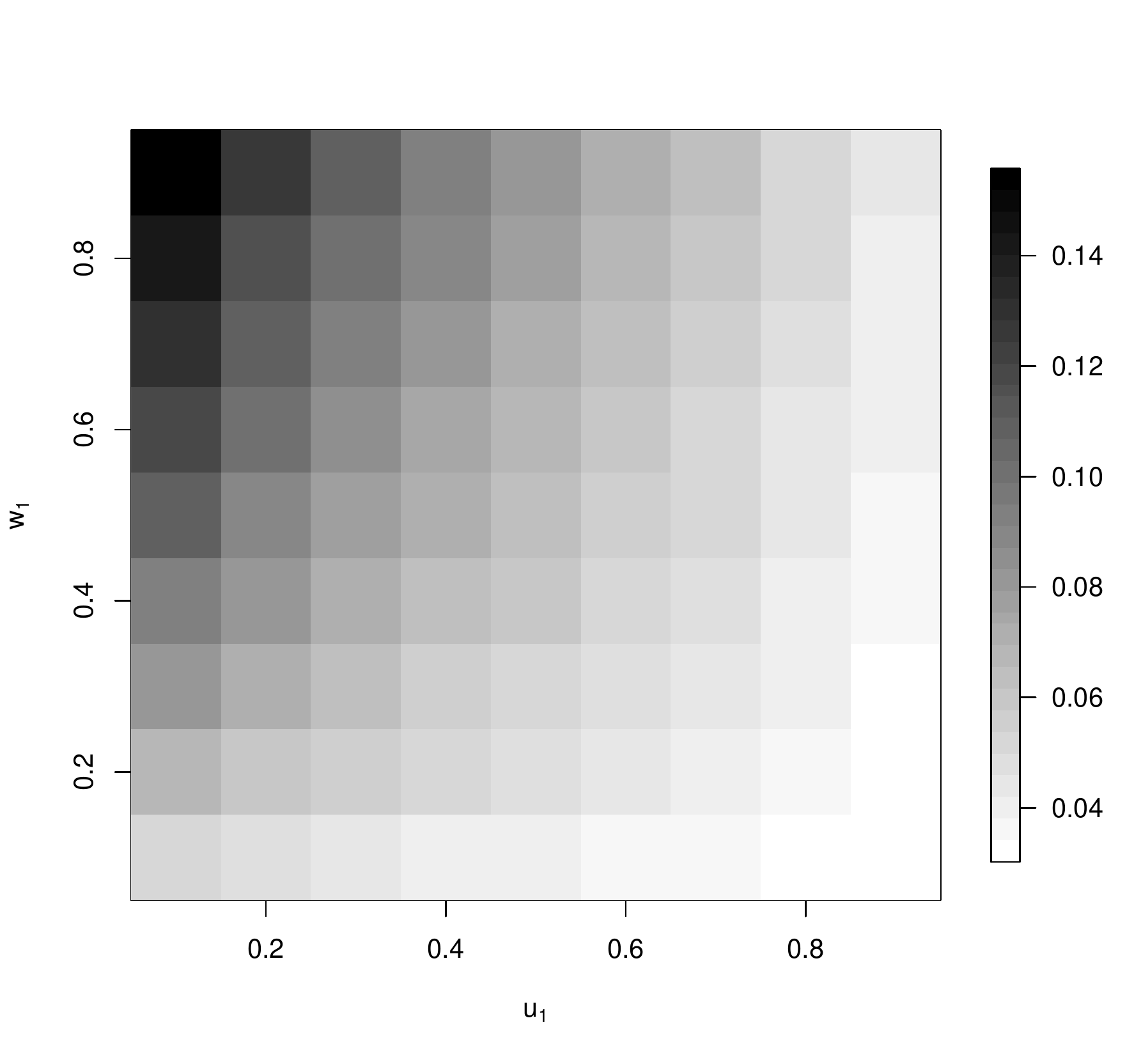}
\end{center}
\caption{\label{figImageInflation}Worst case type I error for various choices of weights and information fractions.}
\end{figure}

The two parameters that govern the size of (\ref{findProb}) are $w_1$ and $u_1$. Larger values of $w_1$ reflect an increased weighting of the  first-stage data, which increases the potential inflation. In addition, a low value for  $u_1$ increases the window of opportunity for stopping on a random high. Figure \ref{figImageInflation} shows that for a nominal $\alpha=0.025$ level test, the worst-case type I error can be up to $15\%$ when $u_1=0.1$ and $w_1=0.9$. As $u_1\rightarrow 0$ the worst-case type I error rate tends to 1 for any value of $w_1$ \citep[see, e.g.,][]{proschan92}.
  
\section{Example \label{secEx}}

The upper bound on the type I error rate, as depicted in Figure \ref{figImageInflation}, varies substantially across $w_1$ and $u_1$.  The following example, simplified from  \cite{irle12}, is intended to give an indication of what can be expected in practice.

 A randomized trial is set up to compare chemotherapy (C) with a combination of radiotherapy and chemotherapy (E). The anticipated median survival time on C is 14 months. If E were to increase the median survival time to 20 months then this would be considered a clinically relevant improvement. Assuming exponential survival times, this gives anticipated hazard rates $\lambda_C=0.050$ and $\lambda_E=0.035$, and a target log hazard ratio of $\theta_R=-\log(\lambda_E/\lambda_C)=0.36$. If the error rates for testing $H_0:\theta=0$ against $H_a:\theta=\theta_R$ are $\alpha=0.025$ (one-sided) and $\beta=0.2$, the required number of deaths (assuming equal allocation) is
\[
d_{1,2}=4\left[\left\lbrace \Phi^{-1}(1-\alpha)+\Phi^{-1}(1-\beta)\right\rbrace/\theta_R\right]^2\approx 248. 
\]
 If $8$ patients per month are recruited at a uniform rate throughout an initial period of $40$ months, and the survival times of these patients are followed-up for an additional $20$ months after the end of this period, then standard sample size formulae \citep[][Section 9.2.3.]{machin97} tell us we can expect to observe around 250 deaths by the time of the final analysis.

Now imagine, as \cite{irle12} did, that an interim look is performed after 60 deaths, observed 23 months after the start of the trial. At this point in time, $190$ patients have been recruited.   Based on the interim results, it is decided to increase the total required number of events from $d_{1,2}$ to $d^*_{1,2}$. 

At the time of the $248$th death, i.e.,  the originally planned study end $T_{1,2}$, suppose we observe that $170$ of these deaths have come from patients recruited prior to the interim look. We have our weights (\ref{eqnIrleWeights}),
\[
w_1=(170/248)^{1/2}\text{ and }w_2=(78/248)^{1/2}.
\] 
At this point we make a note of the standardized first-stage logrank score statistic $S_{1}(T_{1}):=S_1(T_{1,2})$ and hence $p_1$ from (\ref{eqnP1}), and continue to follow-up survival times  until  a total of $d^*_{1,2}$  deaths have been observed. Once these additional deaths have been observed, $p_2$ can be found from (\ref{eqnP2X+}), and combined with $p_1$ to give the adaptive test statistic (\ref{eqnStandardTS}). 

Notice that $w_1=(170/248)^{1/2}$ and, ignoring any potential censoring,  $u_1=D_1(T_1)/D_1(T^{\max}) =170/190$. In this case a naive application of the test statistic (\ref{eqnZStar}) leads to an upper bound on the type I error rate of $0.040$. The inflation is not enormous, owing to the relatively slow recruitment rate, but it is not hard to imagine more worrying scenarios.

Suppose, for example, that the trial design called for $48$ patients to be recruited  per month for $12$ months, with $8$ months of additional follow-up. Further suppose that an interim analysis took place 6 months into the trial, by which time $288$ patients had been recruited, and a decision was made to  increase the total number of events. Given the anticipated $\lambda_C$ and $\lambda_E$, a plausible scenario is that $147$ of the first  $248$ events come from first-stage patients, implying that $w_1=(147/248)^{1/2}$ and  $u_1=147/288$. This gives  an upper bound  (\ref{findProb}) of $0.066$.

\subsection{An alternative level-$\alpha$ test\label{secAlternative}}

A possible rationale for using  (\ref{eqnZStar}), instead of (\ref{eqnAgain}), is that the final test statistic takes into account all available survival times, i.e., does not ignore any data.  If one is  unprepared  to give up the guarantee of type I error control, an alternative test can be found by increasing the cut-off value for $Z^*$ from $\Phi^{-1}(1-\alpha)$ to $k^*$ such that 
\[
\int_{0}^{1}P_{\theta=0}\left[ \max_{u=u_1}^{1} B(u) > u^{1/2}w_1^{-1}\left\lbrace k^*-w_2\Phi^{-1}( 1- p_2)\right\rbrace\right] \ud p_2=\alpha
\]
This will, of course, have a knock on effect on power. Table \ref{tabKStar} gives an impression of how much the cutoff is increased from $1.96$ when $\alpha=0.025$ (one sided).

\begin{table}[ht]\caption{Cutoff values for corrected level-$0.025$ test.\label{tabKStar}}
\centering
\begin{tabular}{rr|rrrrrrrrr}
&&&&&&$u_1$&&&& \\
  
& & 0.1 & 0.2 & 0.3 & 0.4 & 0.5 & 0.6 & 0.7 & 0.8 & 0.9 \\ 
  \hline
&0.1&2.29&2.25&2.21&2.19&2.16&2.13&2.11&2.08&2.04\\
&0.2&2.41&2.35&2.31&2.27&2.23&2.20&2.16&2.12&2.07\\
&0.3&2.50&2.43&2.38&2.34&2.30&2.25&2.21&2.16&2.10\\
&0.4&2.58&2.50&2.44&2.39&2.34&2.30&2.25&2.19&2.12\\
$w_1$&0.5&2.64&2.56&2.49&2.44&2.38&2.33&2.27&2.21&2.14\\
&0.6&2.70&2.60&2.53&2.47&2.42&2.36&2.30&2.23&2.15\\
&0.7&2.74&2.64&2.57&2.51&2.45&2.39&2.33&2.26&2.17\\
&0.8&2.79&2.68&2.60&2.54&2.48&2.41&2.35&2.28&2.18\\
&0.9&2.83&2.72&2.64&2.57&2.50&2.43&2.37&2.29&2.19
\end{tabular}
\end{table}

In assessing the effect on power, at least four probabilities appear relevant:
\begin{enumerate}
\item[\textbf{A}.] $P_{\theta=\theta_R} \left[ 2w_1 S_{1}(T_{1})/\left\lbrace D_{1}(T_1)\right\rbrace^{1/2}+w_2\Phi^{-1}( 1- p_2) > \Phi^{-1}(1-\alpha) \right]$.
\item[\textbf{B}.] $P_{\theta=\theta_R} \left[ 2w_1 S_{1}(T_{1})/\left\lbrace D_{1}(T_1)\right\rbrace^{1/2}+w_2\Phi^{-1}( 1- p_2 ) > k^* \right]$.
\item[\textbf{C}.] $P_{\theta=\theta_R}\left[ 2w_1 S_{1}(T^{\max})/\left\lbrace D_{1}(T^{\max})\right\rbrace^{1/2}+w_2\Phi^{-1}(1-p_2)>k^*\right]$.
\item[\textbf{D}.] $P_{\theta=\theta_R} \left\lbrace \max_{t\geq T_1}Z(t) > k^* \right\rbrace$.
\end{enumerate}
Power definition \textbf{A} corresponds to the ``correct'' adaptive test. $\textbf{B}$ can be thought of as a lower bound on the power of the alternative level-$\alpha$ test, where one conscientiously specifies the increased cutoff value $k^*$ (in anticipation of unpredictable end of first-stage follow-up),  but it then turns out that the trial finishes at the prespecified time point anyhow, i.e., $T^*=T_{1}$. Definition \textbf{C} can be thought of as the power of the alternative test if the trial is always prolonged such that all first-stage events are observed. Definition \textbf{D}, on the other hand, can be interpreted as the power of the alternative level-$\alpha$ test, taken at face value. In other words, assuming that  one takes the opportunity to stop follow-up of first-stage patients when $Z(t)$ is at its maximum. This can be calculated using the same techniques as in Section \ref{secUnspecified}.

\begin{figure}
\centering
\begin{subfigure}{.45\textwidth}
  \centering
  \includegraphics[width=.9\linewidth]{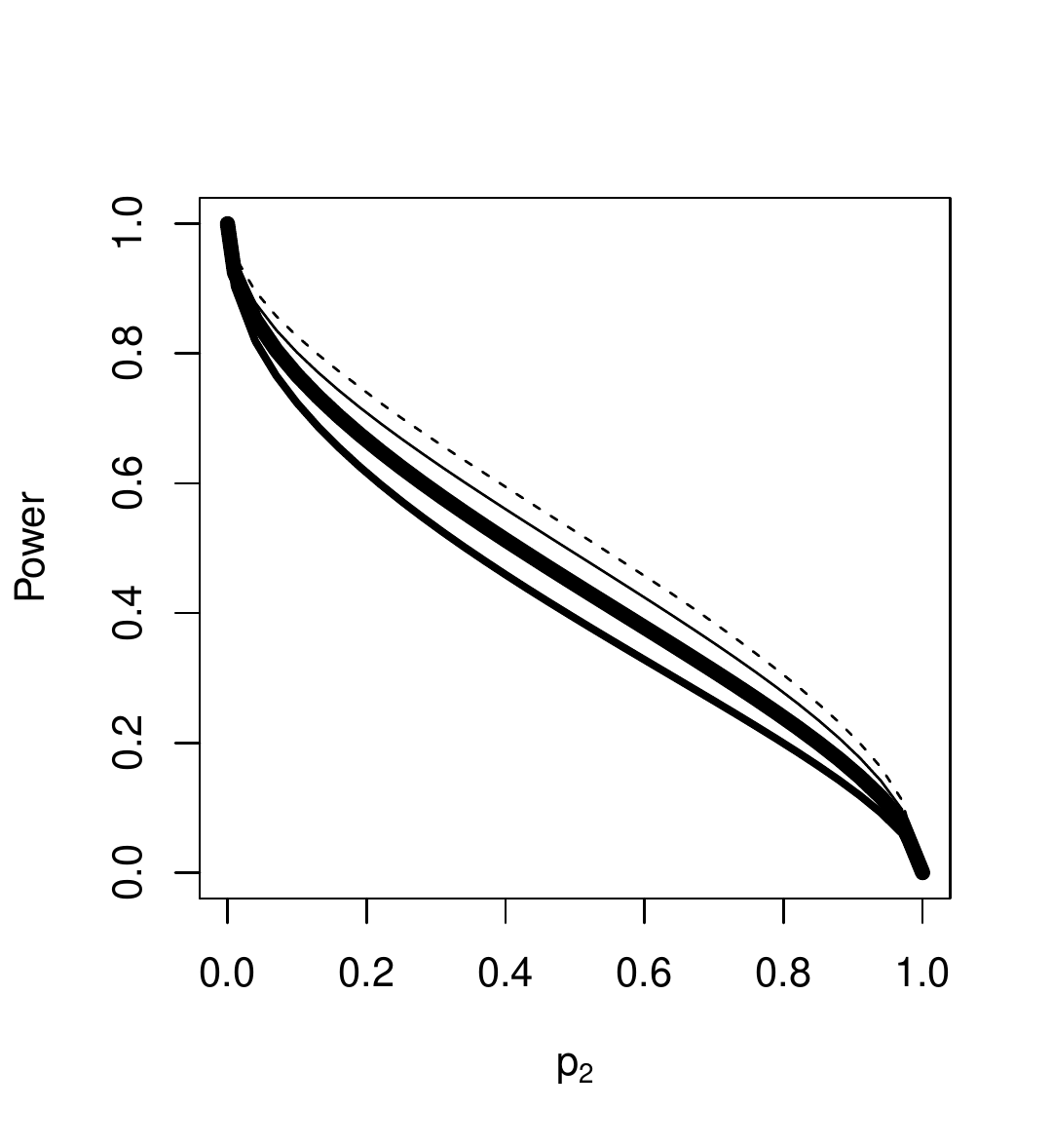}
  \caption{~}
  \label{fig:sub1}
\end{subfigure}
\begin{subfigure}{.45\textwidth}
  \centering
  \includegraphics[width=.9\linewidth]{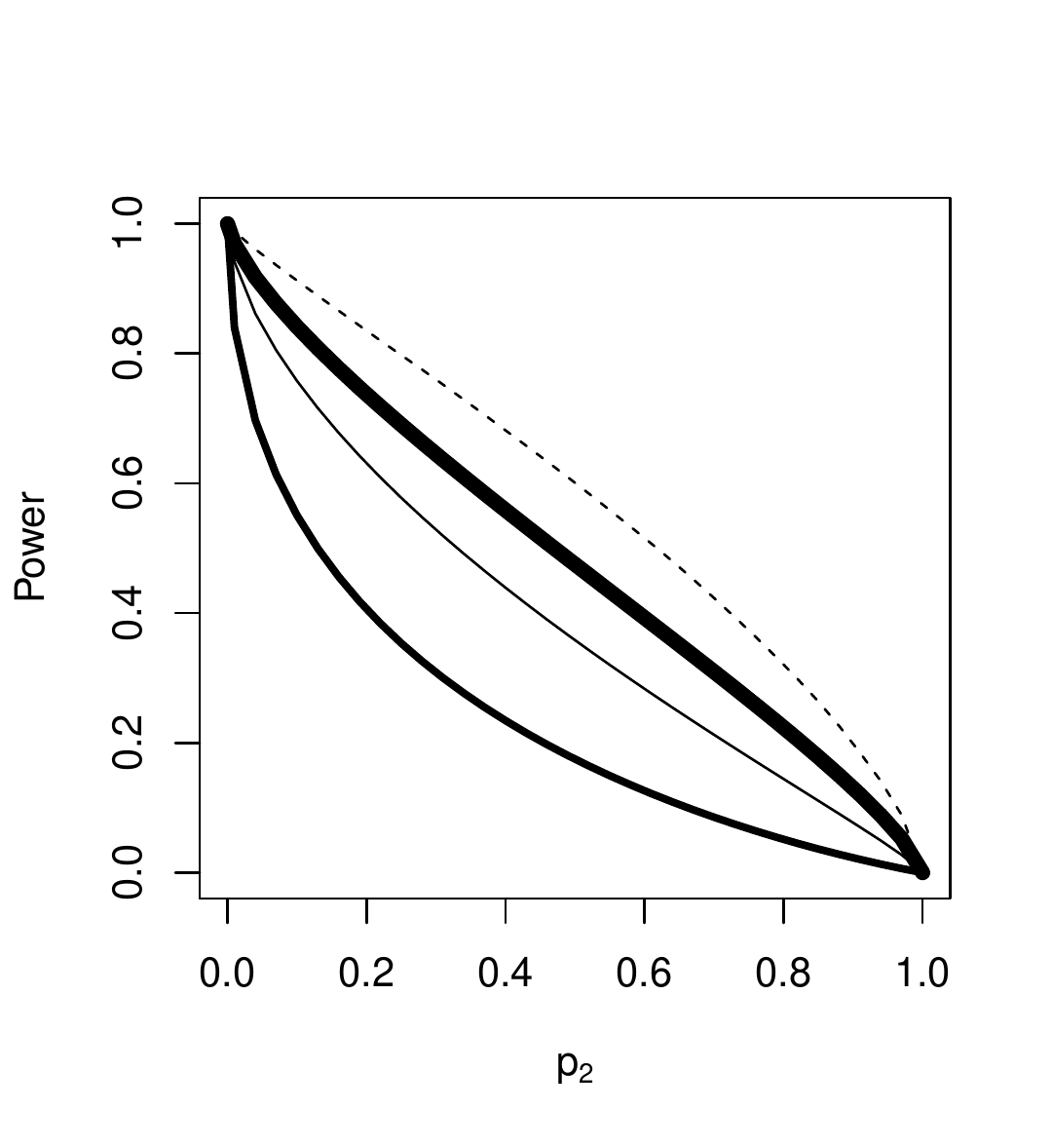}
  \caption{~}
  \label{fig:sub2}
\end{subfigure}
\caption{Conditional power as defined by \textbf{A} (thin line), \textbf{B} (medium line), \textbf{C} (thick line) and \textbf{D}y
 (dashed line) given $p_2$, under the two scenarios described in  Example \ref{secEx}. Scenario (a): $D_{1}(T_1)=170$, $D_{1}(T^{\max})=190$, $w_1=(170/248)^{1/2}$ and $\theta_R=0.36$. Scenario (b): $D_{1}(T_1)=147$, $D_{1}(T^{\max})=288$, $w_1=(147/248)^{1/2}$ and $\theta_R=0.36$.}
\label{figExes}
\end{figure}

Figure \ref{figExes} shows the power of the trial described in Example \ref{secEx}, according to \textbf{A}-\textbf{D}. The power has been evaluated conditional on $p_2$, as this is a random variable common to all four definitions. The increased cutoff value of the alternative level-$\alpha$ test leads to a sizeable loss of power if the trial completes as planned.  In the second scenario at least, the loss of power can be more than made up for when the trial is prolonged. However, if there is an \textit{a-priori} reasonable probability of prolonging the trial, then one could just start with a larger sample size/ required number of events.

In general, the differences between power definitions \textbf{A}-\textbf{D} will tend to follow the same pattern as in Figure \ref{figExes}. The degree to which they differ will depend on $w_1$, $D_{1}(T_1)$, $D_{1}(T^{\max})$ and $\theta_R$. Intuitively, larger $w_1$ and smaller $u_1$ will lead to a greater loss of power going from \textbf{A} to \textbf{B}, but with a greater potential gain in power going from \textbf{B} to \textbf{C} (or \textbf{D}). The actual gain in power from \textbf{B} to \textbf{C} (or \textbf{D}) will be greatest for large values of $\theta_R$

\subsection{Diverging hazard rates \label{secIncreasingH}}

Consider the second trial design in Section \ref{secEx}, where recruitment proceeds at a uniform rate of 48 patients per month for 12 months, with 8 months of additional follow-up. Suppose, however, that the true hazard rates are not proportional. Rather, $h_E(\tau)=0.04$ and $h_C^{-1}(\tau)=0.04^{-1}-0.6\tau$ for $\tau \in \left(0, 30\right)$, where $\tau$ denotes the time in calendar months since randomization. Simulating a realization of this trial, $295$ patients are recruited in the first six months, by which time there have been  $18$ deaths on $C$, and $15$ deaths on $E$. Suppose that at this point it is decided to increase the target number of events from $d_{1,2}=248$ to $d_{1,2}^*= 350$. At time $T_{1,2}$, the number of deaths from first-stage patients -- those patients recruited in the first six months -- is $D_1(T_{1,2})= 151$, such that $w_1=(151/248)^{1/2}$, $u_1= 151/ 295$ and $k^*= 2.41 $. The logrank score statistics based on first-stage patients is $S_{1}(T_{1,2})=7.6$, giving a first-stage p-value (\ref{eqnP1}) of
\begin{equation*}
p_1  =  1 - \Phi \left\lbrace 2(7.6)/151^{1/2}\right\rbrace = 0.108
\end{equation*}
and a conditional error probability (\ref{eqnVarPhi}) of $E_{H_0}\left\lbrace \varphi \mid S_1(T_{1,2}) = 7.6 \right\rbrace = 0.213$, using (\ref{eqnDist}).

\begin{figure}
\centering
\begin{subfigure}{.45\textwidth}
  \centering
  \includegraphics[width=.9\linewidth]{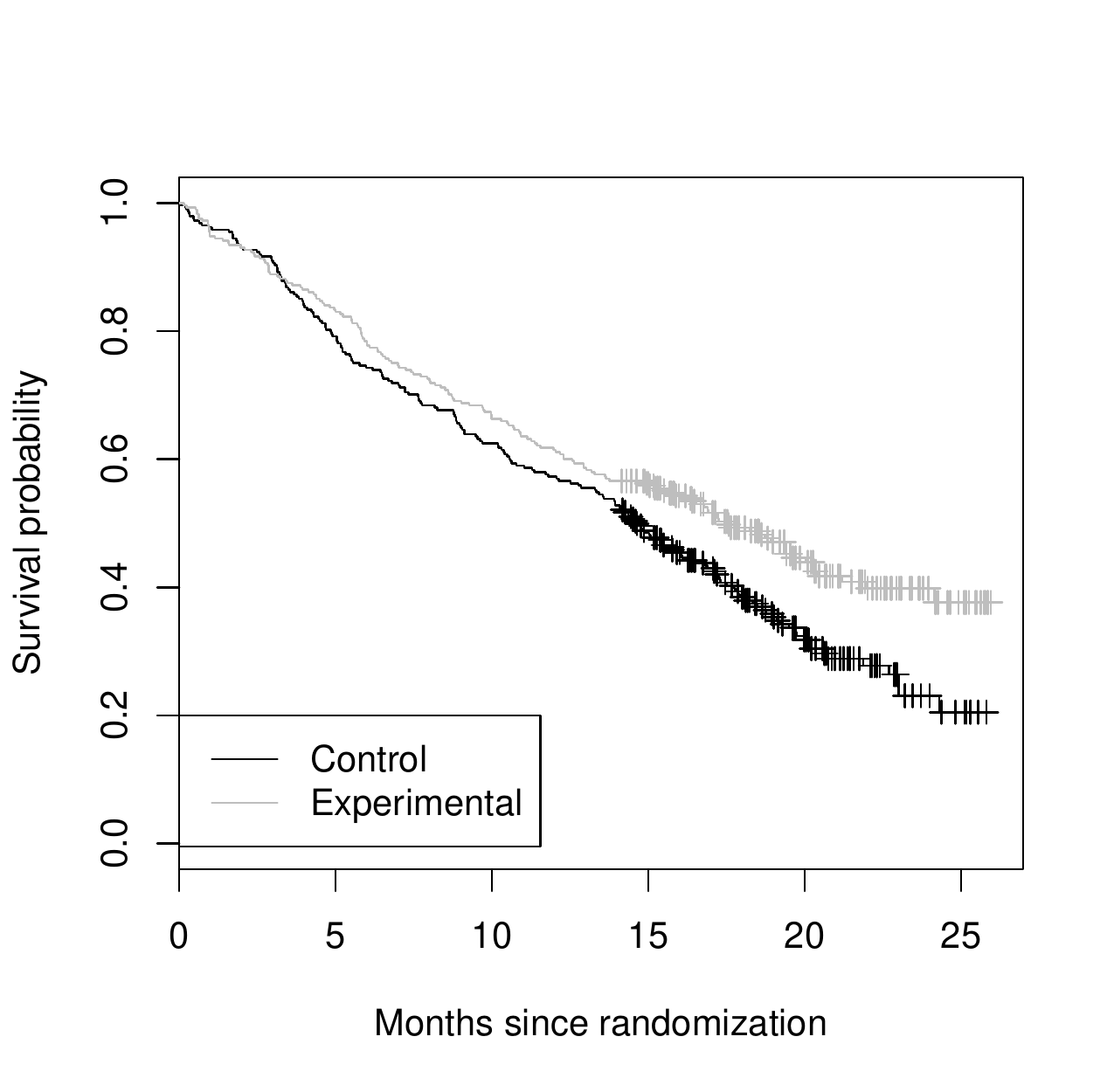}
  \caption{~}
  \label{fig:sub1}
\end{subfigure}
\begin{subfigure}{.45\textwidth}
  \centering
  \includegraphics[width=.9\linewidth]{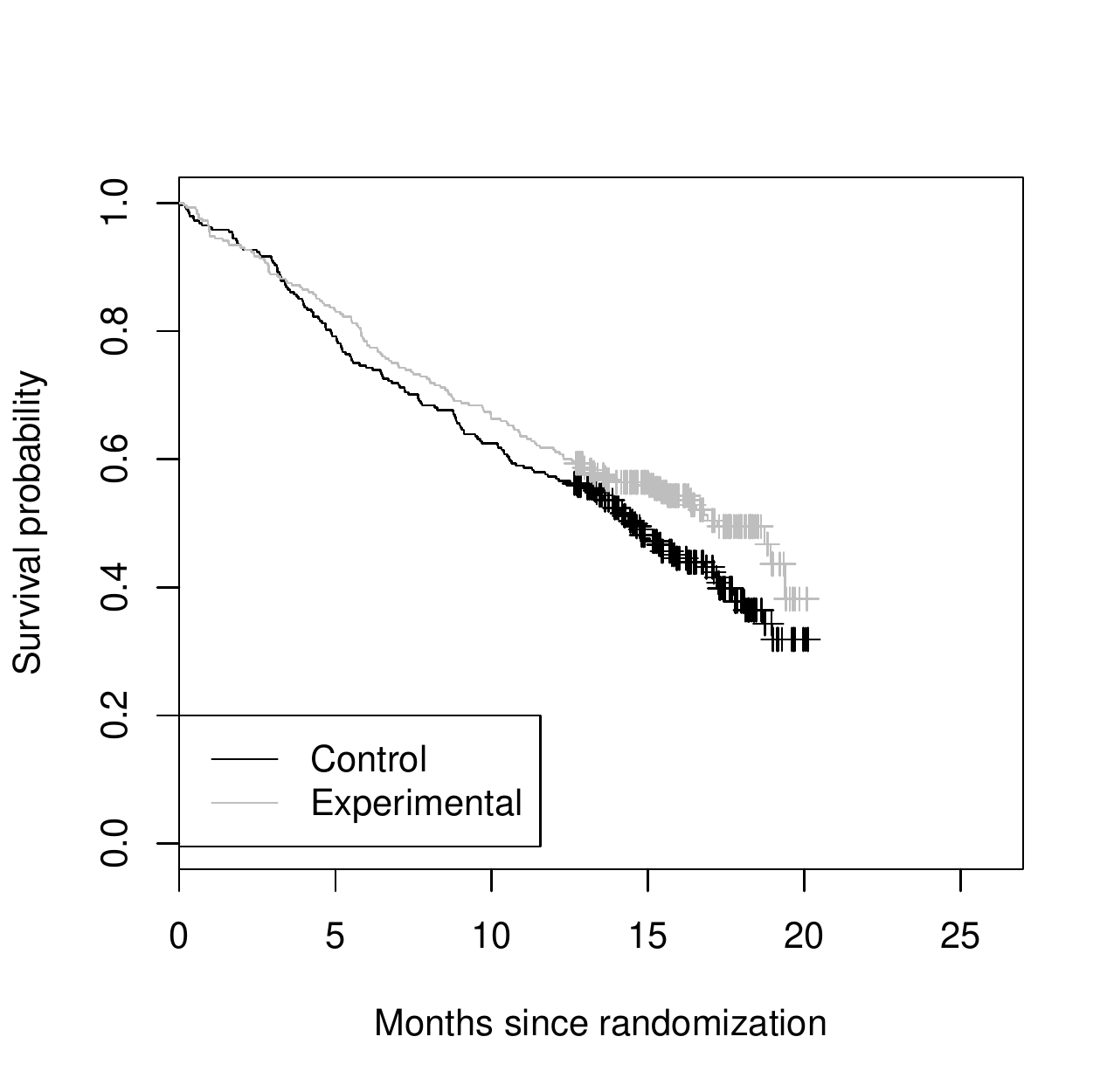}
  \caption{~}
  \label{fig:sub2}
\end{subfigure}
\caption{Kaplan-Meier plots corresponding to the hypothetical trial described in Section \ref{secIncreasingH} based on (a) all available survival times at $T_{1,2}^*$, and (b) with first-stage and second-stage patients censored at $T_{1,2}$ and $T_{1,2}^*$, respectively.}
\label{figKM}
\end{figure}

The survival data at time $T_{1,2}^*$,  occurring approximately $26$ months into the trial, is plotted in Figure \ref{figKM}. On the left-hand-side, all survival times have been included in the Kaplan-Meier curves. There is an obvious divergence in the survival probabilities on the two treatments. However, the test decision of \cite{irle12} may only use the data as depicted on the right-hand-side, where the survival times of first-stage patients have been censored at time $T_{1,2}$. They are liable to reach an inappropriate conclusion. In this case, $199$ out of the first $350$ events are from patients recruited in the first six months, and the logrank score statistic based on first-stage patients is $S_1(T_{1,2}^*)=16$. The new cutoff value $b^*$ must be found to solve 
\begin{equation*}
E_{H_0}\left\lbrace \psi_{x_1^{\text{int}}}\mid S_1(T_{1,2}^*) = 16\right\rbrace  =  P_{H_0}\left\lbrace 2 S_{1,2}(T_{1,2}^*)/350^{1/2} \geq b^* \mid S_1(T_{1,2}^*) = 16\right\rbrace = 0.213,
\end{equation*}
which gives $b^* = 2.76$, using (\ref{eqnDist}).  The logrank statistic based on all survival times at $T_{1,2}^*$ is $S_{1,2}(T_{1,2}^*) =  25$ and the test decision is
\[
\psi_{x_1^{\text{int}}}= \mathbf{1}\left\lbrace 2(25)/350^{1/2} \geq 2.76 \right\rbrace = 0,
\]
i.e., one cannot  reject the null hypothesis. As shown in Section \ref{secIrle}, the same decision could have been reached by finding the second-stage p-value (\ref{eqnP2X+})
\[
p_2 = 1- \Phi\left[ 2 \left\lbrace 25-16\right\rbrace / (350 - 199)^{1/2}\right] = 0.071,
\]
computing the adaptive test statistic (\ref{eqnStandardTS}),
\begin{equation*}
Z  =  w_1\Phi^{-1}(1-p_1)+w_2\Phi^{-1}(1-p_2)  =   1.88,
\end{equation*}
and comparing with $\Phi^{-1}(1-\alpha)\approx 1.96$. The number of events that have been ignored in making this decision is $D_{1}(T_{1,2}^*)-D_1(T_{1,2}) = 48$.

If, on the other hand, one had prespecified the alternative test of Section \ref{secAlternative}, then one would be permitted to replace $\Phi^{-1}(1-p_1)$ in the adaptive test statistic with the value of the standardized logrank statistic at time $T_{1,2}^*$. In this case one would be able to reject the null hypothesis, as
\begin{equation*}
Z(T_{1,2}^*)  =  2w_1S_1(T_{1,2}^*)/D_1(T_{1,2}^*)^{1/2} + w_2 \Phi^{-1}(1-p_2) =  2.69  >  k^*.
\end{equation*}

\section{Discussion \label{secDiscussion}}

Adaptive design methodology -- developed over the past two decades to cope with mid-study protocol changes in confirmatory clinical trials -- is becoming increasingly accepted by regulatory agencies \citep{elsaesser13}. It is therefore unfortunate that the important case of time-to-event data is not easily handled by the standard theory. As far as survival data are concerned, all proposed solutions have limitations and there is a trade-off between strict type I error control, power, flexibility, the use of all interim data to substantiate interim decision making, and the use of all available data in making the test decision at the final analysis.

The proposed solutions of \cite{irle12} and \cite{jenkins10} offer strict type I error control and allow full use of the interim data. The \cite{jenkins10} method allows one to change the recruitment rate at the interim analysis -- something that is disallowed by \cite{irle12}, where one is only permitted to increase the observation time. On the other hand, \cite{irle12} is more flexible in the sense that the timing of the interim analysis need not be prespecified.  In both cases, the final test decision only depends on a subset of the recorded survival times, i.e., part of the observed data is ignored. This is usually deemed unacceptable by regulators. Furthermore, it is the long-term data of patients recruited prior to the interim analysis that is ignored, such that more emphasis is put on early events in the final decision making. This neglect becomes egregious when there is specific interest in learning about the long-term parts of the survival curves.

We have therefore proposed an alternative procedure which offers the same type I error control and flexibility as \cite{jenkins10} and \cite{irle12}, in addition to a final test statistic that takes into account all available survival times. However, in order to achieve this, a worst-case adjustment is made a-priori in the planning phase. If no design modifications are performed at the interim analysis, the worst-case critical boundary must nevertheless be applied. This results in a loss of power.

Methods based on the independent increments assumption have been only briefly mentioned in Section \ref{secII}. They suffer from the limitation that decision makers must be blinded  to short-term data at the interim analysis. On the other hand, subject to this blinding being imposed, the type I error rate is controlled and the final test decision is based on all available survival times. This could therefore be a viable option in situations where the short-term data is sparse or relatively uninformative. Yet another option, if one is prepared to give up strict type I error control, is simply to  use the usual logrank test at the final analysis. The true operating characteristics of such a procedure are unclear, owing to the  complex dependence on the interim data.

Our alternative level-$\alpha$ test may have practical applications in multi-arm survival trials \citep{jaki2013considerations} and  adaptive enrichment designs. In this case, one must take great care in applying the methodology of \cite{jenkins10} or \cite{irle12}. One specific issue is that dropping a treatment arm will affect recruitment rates on other arms. Also, for treatment regimes that have to be given continuously over a period of time, it would be unethical to keep treating patients on treatment arms that have been dropped for futility. This may affect the timing of analyses on other arms. Incorporating some flexibility into the end of patient follow-up could confer advantages here.  More research is 
 needed in this area.


The usefulness of performing design modifications has to be thoroughly assessed on a case-by-case basis in the planning phase. Interim data may be highly variable, and the interim survival results may be driven mainly by early events. Consequently, the interim data may be too premature to allow a sensible interpretation of the whole survival curves and may not be a reliable basis for adaptations.

In this respect, the best advice might be to thoroughly assess the characteristics of adaptive trial designs in comparison with more standard approaches, and to plan for  adaptations only in settings where the advantages are compelling. If in the planning phase there is a strong likelihood that the number of patients will need to be increased, or the observation time extended, our analysis has shown that there is no uniformly best design. All proposals to implement adaptive survival designs have their limitations. If the main objective is strict type I error control when using all data, then our proposal should be considered as a valid option.

\section*{Appendix A}

\subsection*{Connection between conditional error and combination test}

The cut-off $b^*$ satisfies
\begin{IEEEeqnarray*}{rCl}
E_{H_0}\left\lbrace \varphi\mid S_1(T_{1,2})=s_1\right\rbrace &=& P_{H_0}\left\lbrace 2 S_{1,2}(T_{1,2}^*) / (d_{1,2}^*)^{1/2}\geq b^* \mid S_1(T_{1,2}^*)=s_1^*\right\rbrace\\
&=& P_{H_0}\left[2\left\lbrace S_{1,2}(T_{1,2}^*)-S_{1}(T_{1,2}^*)\right\rbrace / \left\lbrace d_{1,2}^*-D_{1}(T_{1,2}^*)\right\rbrace^{1/2}\geq c^* \mid S_1(T_{1,2}^*)=s_1^* \right],
\end{IEEEeqnarray*}
which implies that $c^*=\Phi^{-1}\left[1-E_{H_0}\left\lbrace \varphi\mid S_1(T_{1,2})=s_1 \right\rbrace\right]$, using (\ref{eqnDist}). Therefore,
\begin{IEEEeqnarray*}{rCl}
\psi_{X_1^{\text{int}}}=1 &\Leftrightarrow & 2 S_{1,2}(T_{1,2}^*) / (d_{1,2}^*)^{1/2}\geq b^*\\
 &\Leftrightarrow & 2\left\lbrace S_{1,2}(T_{1,2}^*)-S_{1}(T_{1,2}^*)\right\rbrace / \left\lbrace d_{1,2}^*-D_{1}(T_{1,2}^*)\right\rbrace^{1/2}\geq c^* \\
 &\Leftrightarrow & \Phi^{-1}(1-p_2) \geq \Phi^{-1}\left[1-E_{H_0}\left\lbrace \varphi\mid S_1(T_{1,2})=s_1\right\rbrace\right]\\
 &\Leftrightarrow & p_2 \leq  E_{H_0}\left\lbrace \varphi\mid S_1(T_{1,2})=s_1\right\rbrace.
\end{IEEEeqnarray*}

The conditional error probability, $E_{H_0}\left\lbrace \varphi\mid S_1(T_{1,2})=s_1\right\rbrace$, can be found from the joint distribution (\ref{eqnDist}) at calendar time $T_{1,2}$. Omitting the argument $T_{1,2}$ from $S_{1}$, $S_{1,2}$, $D_{1}$ and $D_{1,2}$:
\small
\begin{IEEEeqnarray*}{rCl}\label{eqnCE/CF}
E_{H_0}\left\lbrace \varphi\mid S_1=s_1\right\rbrace&=& P_{H_0}\left\lbrace 2 S_{1,2}/(D_{1,2})^{1/2}>\Phi^{-1}(1-\alpha)\mid S_{1} =s_1\right\rbrace \\
 & = & P_{H_0}\left[ 2(S_{1,2}-S_{1})/(D_{1,2}-D_{1})^{1/2}>\Phi^{-1}(1-\alpha)\left\lbrace D_{1,2}/(D_{1,2}-D_{1})\right\rbrace^{1/2} \right. \\
 & & \hfill\left. -2S_{1}/(D_{1,2}-D_{1})^{1/2} \mid S_{1}=s_1 \right]\\
 & = & 1-\Phi\left[ \Phi^{-1}(1-\alpha)\left\lbrace D_{1,2}/(D_{1,2}-D_{1})\right\rbrace^{1/2}- \Phi^{-1}(1-p_1)\left\lbrace D_{1}/(D_{1,2}-D_{1})\right\rbrace^{1/2} \right]
\end{IEEEeqnarray*}
\normalsize
and therefore $p_2 \leq E_{H_0}\left\lbrace \varphi\mid S_1(T_{1,2})=s_1\right\rbrace$  if and only if 
 \[
\left\lbrace D_{1}(T_{1,2})/d_{1,2}\right\rbrace^{1/2}\Phi^{-1}(1-p_1)+\left[\left\lbrace d_{1,2}-D_{1}(T_{1,2})\right\rbrace/d_{1,2}\right]^{1/2}\Phi^{-1}(1-p_2) \geq \Phi^{-1}(1-\alpha).
\]

\section*{Appendix B}

\subsection*{Computation of (\ref{findProb})}

For simplicity, consider replacing the square root boundary in (\ref{eqnWienWeights}) with a linear boundary. Conditional on $p_2$, our problem is to find $P_{\theta=0}\left\lbrace B(u) < au+b, ~u_1<u\leq 1 \right\rbrace$, where $a$ and $b$ are found by drawing a line through
\[
u_1,~u_1^{1/2}w_1^{-1}\left\lbrace \Phi^{-1}(1-\alpha)-w_2\Phi^{-1}(1-p_2)\right\rbrace 
\]
and
\[
 1,~w_1^{-1}\left\lbrace \Phi^{-1}(1-\alpha)-w_2\Phi^{-1}(1-p_2)\right\rbrace.
\]

For constants $a$, $b$ and $c$, with $b,c> 0$,  \cite{siegmund86} shows that
\begin{equation*}\label{eqnSiegmund}
P_{\theta=0}\left\lbrace B(u) \geq au + b, \text{for some } 0<u \leq c \mid W(c)=x\right\rbrace = \exp\left\lbrace - 2b(ac+b-x)/c\right\rbrace
\end{equation*}
and integrating  over $x$ gives
\begin{equation}\label{eqn0toT}
P_{\theta=0}\left\lbrace B(u) < au+b,~ u\leq c \right\rbrace= \Phi\left\lbrace (ac+b)/c^{1/2}\right\rbrace - \exp(-2ab)\Phi\left\lbrace (ac-b)/c^{1/2}\right\rbrace.
\end{equation}
Therefore, conditioning on the value of $B(u_1)$,
\begin{IEEEeqnarray*}{rCl}
P_{\theta=0}\left\lbrace B(u) < au+b,~ u_1<u\leq 1 \right\rbrace & = & \int_{-\infty}^{au_1} P_{\theta=0}\left\lbrace B(u) < au+b , ~u_1 < u\leq 1 \mid B(u_1)=x \right\rbrace\ud P_{u_1}(x;0)\\
  & = &\int_{-\infty}^{au_1} P_{\theta=0}\left\lbrace B(u+u_1)-x  \right.\\
 &  & \hfill\left.< a(u+u_1)+b -x,~ 0< u\leq 1-u_1 \mid B(u_1)=x \right\rbrace\ud P_{u_1}(x;0) \\
  & = &\int_{-\infty}^{au_1} P_{\theta=0}\left\lbrace B(v) < av +au_1 +b- x , ~0 < v\leq 1-u_1 \right\rbrace\ud P_{u_1}(x;0)\\
  & = &\int_{-\infty}^{au_1} \Phi\left\lbrace \frac{a+b-x}{(1-u_1)^{1/2}}\right\rbrace \\
 & &\hfill- \exp\left\lbrace -2a(au_1+b-x)\right\rbrace\Phi \left\lbrace \frac{a(1-2u_1)-b+x}{(1-u_1)^{1/2}}\right\rbrace\ud P_{u_1}(x;0).
\end{IEEEeqnarray*}
Greater accuracy can be achieved by replacing the square root boundary with a piece-wise linear boundary, in which case one must condition on the value of the Brownian motion at each of the cut-points \citep{wang97}. 
\bibliographystyle{apalike}
\bibliography{multcomplit3}

\end{document}